\documentstyle[12pt]{article}

\catcode`\@=11
\def\marginnote#1{}
\newcount\hour
\newcount\minute
\newtoks\amorpm
\hour=\time\divide\hour by60
\minute=\time{\multiply\hour by60 \global\advance\minute
by-\hour}\edef\standardtime{{\ifnum\hour<12
\global\amorpm={am}%
        \else\global\amorpm={pm}\advance\hour by-12 \fi
        \ifnum\hour=0 \hour=12 \fi
        \number\hour:\ifnum\minute<10
0\fi\number\minute\the\amorpm}}
\edef\militarytime{\number\hour:\ifnum\minute<10
0\fi\number\minute}

\def\draftlabel#1{{\@bsphack\if@filesw {\let\thepage\relax
   \xdef\@gtempa{\write\@auxout{\string
      \newlabel{#1}{{\@currentlabel}{\thepage}}}}}\@gtempa
   \if@nobreak \ifvmode\nobreak\fi\fi\fi\@esphack}
        \gdef\@eqnlabel{#1}}
\def\@eqnlabel{}
\def\@vacuum{}
\def\draftmarginnote#1{\marginpar{\raggedright\scriptsize\tt#1}}
\def\draft{\oddsidemargin -.5truein
        \def\@oddfoot{\sl preliminary draft \hfil
        \rm\thepage\hfil\sl\today\quad\militarytime}
        \let\@evenfoot\@oddfoot \overfullrule 3pt
        \let\label=\draftlabel
        \let\marginnote=\draftmarginnote

\def\@eqnnum{(\theequation)\rlap{\kern\marginparsep\tt\@eqnlabel}%
\global\let\@eqnlabel\@vacuum}  }

\def\numberbysection{\@addtoreset{equation}{section}
        \def\theequation{\thesection.\arabic{equation}}}

\def\underline#1{\relax\ifmmode\@@underline#1\else
 $\@@underline{\hbox{#1}}$\relax\fi}

\catcode`@=12
\relax

\numberbysection

\advance \topmargin by -\headheight
\advance \topmargin by -\headsep

\evensidemargin \oddsidemargin
\def\br{\begin{eqnarray}}
\def\er{\end{eqnarray}}
\def\be{\begin{equation}}
\def\ee{\end{equation}}

\def\({\left(}
\def\){\right)}

\relax

\def\a{\alpha}

\def\b{\beta}

\def\d{\delta}

\def\g{\gamma}
\def\G{\Gamma}

\def\l{\lambda}

\def\pa{\partial}

\def\s{\sigma}

\def\t{\tau}

\def\tp0{\Theta_{+}^{(0)}}
\def\tm0{\Theta_{-}^{(0)}}

\def\vp{\varphi}

\def\f#1#2#3 {f^{#1#2}_{#3}}

\def\win1{{\sf w_{1+\infty}}}

\def\Win1{{\sf W_{1+\infty}}}

\def\rlx{\relax\leavevmode}
\def\inbar{\vrule height1.5ex width.4pt depth0pt}
\def\IZ{\rlx\hbox{\sf Z\kern-.4em Z}}
\def\IR{\rlx\hbox{\rm I\kern-.18em R}}
\def\IC{\rlx\hbox{\,$\inbar\kern-.3em{\rm C}$}}
\def\IN{\rlx\hbox{\rm I\kern-.18em N}}
\def\IO{\rlx\hbox{\,$\inbar\kern-.3em{\rm O}$}}
\def\IP{\rlx\hbox{\rm I\kern-.18em P}}
\def\IQ{\rlx\hbox{\,$\inbar\kern-.3em{\rm Q}$}}
\def\IF{\rlx\hbox{\rm I\kern-.18em F}}
\def\IG{\rlx\hbox{\,$\inbar\kern-.3em{\rm G}$}}
\def\IH{\rlx\hbox{\rm I\kern-.18em H}}
\def\II{\rlx\hbox{\rm I\kern-.18em I}}
\def\IK{\rlx\hbox{\rm I\kern-.18em K}}
\def\IL{\rlx\hbox{\rm I\kern-.18em L}}
\def\one{\hbox{{1}\kern-.25em\hbox{l}}}
\def\0#1{\relax\ifmmode\mathaccent"7017{#1}%
B        \else\accent23#1\relax\fi}

\def\PRL#1#2#3{{\sl Phys. Rev. Lett.} {\bf#1} (#2) #3}
\def\PR#1#2#3{{\sl Phys. Rept.} {\bf#1} (#2) #3}

\def\NPB#1#2#3{{\sl Nucl. Phys.} {\bf B#1} (#2) #3}

\def\PRD#1#2#3{{\sl Phys. Rev.} {\bf D#1} (#2) #3}
\def\PRv#1#2#3{{\sl Phys. Rev.} {\bf #1} (#2) #3}
\def\PLA#1#2#3{{\sl Phys. Lett.} {\bf #1A} (#2) #3}
\def\PLB#1#2#3{{\sl Phys. Lett.} {\bf #1B} (#2) #3}
\def\JMP#1#2#3{{\sl J. Math. Phys.} {\bf #1} (#2) #3}
\def\PTP#1#2#3{{\sl Prog. Theor. Phys.} {\bf #1} (#2) #3}

\def\AoP#1#2#3{{\sl Ann. of Phys.} {\bf #1} (#2) #3}

\def\RMP#1#2#3{{\sl Rev. Mod. Phys.} {\bf #1} (#2) #3}

\def\IJMPD#1#2#3{{\sl Int. J. Mod. Phys.} {\bf D#1} (#2) #3}

\def\JPA#1#2#3{{\sl J. Physics} {\bf A#1} (#2) #3}

\def\MPLA#1#2#3{{\sl Mod. Phys. Lett.} {\bf A#1} (#2) #3}

\def\GRG#1#2#3{{\sl Gen. Rel. Grav.}  {\bf #1} (#2) #3}
\def\NCB#1#2#3{{\sl Nuovo. Cim. }  {\bf B#1} (#2) #3}
\def\NT#1#2#3{{\sl Nature}  {\bf #1} (#2) #3}
\def\AnP#1#2#3{{\sl Annalen Phys}  {\bf #1} (#2) #3}
\def\IJTP#1#2#3{{\sl  Int. J. Theor. Phys.\,}{\bf #1} (#2) #3}
\def\RPP#1#2#3{{\sl  Rept. Prog. Phys.\,}{\bf #1} (#2) #3}
\def\FP#1#2#3{{\sl  Found. Phys.\,}{\bf #1} (#2) #3}

\begin{document}

\begin{center}
{\Large {\bf Metric-scalar gravity with torsion and the
measurability of the  non-minimal coupling}}
\end{center}

\normalsize
\vskip .4in
 \begin{center}

                A. Accioly$^{a}$ and  H. Blas$^{b}$\\

                \vspace{.6 cm}
                $^{a}$ Instituto de F\'\i sica Te\'orica - IFT/UNESP\\
Rua Pamplona 145\\
01405-900  S\~ao Paulo-SP, BRAZIL\\and\\
$^{a}$ Centro Brasileiro de Pesquisas F\'{\i}sicas - CBPF\\
Rua  Dr. Xavier Sigaud 150, 22290-180\\
Rio de Janeiro, RJ, Brazil \\

${}^{b}$ Departamento de Matem\'atica - ICET\\
Universidade Federal de Mato Grosso\\
 Av. Fernando Correa, s/n, Coxip\'o \\
78060-900, Cuiab\'a - MT - BRAZIL\\
  \end{center}

\begin{center}
{\bf ABSTRACT}\\
\end{center}
\par \vskip .3in \noindent

The ``measurability" of the non-minimal coupling is discussed by
considering the correction to the Newtonian static potential in
the semi-classical approach.
 The coefficient of the ``gravitational Darwin term" (GDT) gets redefined by the non-minimal torsion scalar couplings.
 Based on a similar analysis of the GDT in the effective field theory approach to non-minimal scalar we
 conclude that for reasonable values of the couplings the
correction is very small.
\par \vskip 1.7in \noindent
PACS: 04.20.Cv, 04.80.Cc, 11.15.Kc,04.90.+e\\
Keywords: Foldy-Wouthuysen transformation; non-minimal coupling;
torsion; metric-scalar gravity.

\newpage
\section{Introduction}

The interface of (classical) gravitational and quantum realms
\cite{Ahluwalia, kiefer} has been the subject of a considerable
literature for several decades that we can say that the
gravitational effects on quantum  systems  are no more beyond our
reach (see e.g. \cite{Colella, Bonse, Nesvizhevsky}). The
theoretical analysis consisted basically in inserting the
Newtonian gravitational potential into the Sch\"rodinger equation.
On the other hand, considering gravity as an affective field
theory certain quantum predictions can be made (see
\cite{Donoghue, donoghue1} and references therein). One direction
to improve the analysis may be to learn how to handle relativistic
field equations in a curved background space-time and study the
low enough energy and small curvature regions.

A relativistic quantum mechanical system and the effects of
external fields coupled to it can be studied by constructing the
Foldy-Wouthuysen transformation (FWT) \cite{foldy, foldy1}. The
main advantage of the FWT transformation is that the Hamiltonian
and all operators in this representation are block-diagonal. This
transformation holds only in the one-particle approximation where
the loop calculations are not taken into account and this
description is acceptable if the external fields are very weak and
the particle production processes can be neglected. However, there
are very few known problems in flat space that admit an exact FWT
\cite{silenko, nikitin}. In curved space the known exact FWT are
those related to Dirac \cite{obukhov} and spin zero particles
\cite{prd, mpla} coupled to a static spacetime metric. For a
scalar coupled to higher derivative gravity see \cite{ijtp}.

The coupling between the curvature and scalar field of the form
$\lambda\, R\, \phi $  is the only possible local term with
dimensionless coupling constant $\lambda$ \cite{birrel, fulling}.
The values $\lambda=0$ (minimal coupling) and $\lambda=1/6$ (for
massless scalars) are used in the literature. A general value of
$\lambda\neq 0$ is the so-called non-minimal coupling and the
question of which value(s) of $\lambda$ should constitute the
correct coupling to gravity depends on the particular field theory
used for the scalar field (see, e.g. \cite{faraoni1} and
references therein). Given the current theoretical situation it
seems more of an experimental problem to identify which would be
the correct $\lambda$ coupling(s) for the various scalar
particles. It has been suggested that the action of the
gravity-scalar theory should contain, along with the
Einstein-Hilbert action, some non-minimal couplings of the scalar
field with the curvature tensor, see e.g.  \cite{faraoni}.

As for alternatives to classical General Relativity, there are
different options (see e.g. \cite{hehl, hehl1, HammondRPP,
shapiro, Arcos}). Here we consider the so-called Einstein-Cartan
theory \cite{shapiro}. In this context, the torsion field does not
interact minimally with scalar fields, but a non-minimal
formulation allows the introduction of certain interaction terms
with some coupling constants $\xi_{i}$. Moreover, this type of
scalar-torsion interaction is a necessary condition for the
renormalizability of the quantum field theory in curved space-time
with torsion \cite{shapiro, helayel}.

Here we extend the results of \cite{prd, mpla} to the case of a
massive scalar coupled to gravity and torsion. We describe the
formulation of the metric-scalar gravity with torsion and provide
an on-shell action obtained after the substitution of the torsion
fields into the original action by using their relevant field
equations. We will show that the problem of finding the exact FWT
for the metric-scalar gravity with torsion reduces to that of the
non-minimal metric-scalar system and that the torsion field
effects are encoded only in the $\xi_{i}$ dependence of the
redefined non-minimal coupling $\hat{\lambda}$.

 The torsion field effects on the Dirac equation in the
non-relativistic limit have been considered using the
Foldy-Wouthuysen transformation and the semi-classical approach
(see e.g., \cite{Lammerzahl, ryder, Adak, Hammond}).

In the next section we describe the metric-scalar gravity with
torsion and write the on-shell action written in terms only of the
scalar field once the torsion field components are substituted
into the action by making use of their equations of motion. In
section \ref{sec:efw} we write the FWT and the quasi-relativistic
Hamiltonian for the system. In section \ref{sec:hamiltonian} the
Foldy's  Hamiltonian is written for a static background metric and
its properties are discussed.

\section{The metric-scalar gravity with torsion}
\label{sec:torsion} Supposing that the affine connection
$\widetilde{\G}_{\,\b\g}^{\a}$ is not symmetric, i.e., \br
\label{torsion}
\widetilde{\G}_{\,\b\g}^{\a}-\widetilde{\G}_{\,\g\b}^{\a}=
T_{\,\b\g}^{\a}, \er one defines the tensor $T_{\,\b\g}^{\a}$
called torsion.

For our purposes it is useful to decompose the torsion into its
three irreducible components: i) the vector $T_{\b}=
T_{\,\b\a}^{\a}$,  ii) the axial vector  $ S^{\nu}=
\epsilon^{\a\b\mu\nu} T_{\a\b\mu}$, and iii) the tensor
$q_{\a\b\g}$ satisfying $q_{\,\b\a}^{\a}=0$ and
$\epsilon^{\a\b\mu\nu} q_{\a\b\mu}=0$. Then the torsion becomes
\br \label{decom} T_{\a, \b\g}= \frac{1}{3} (T_{\b}g_{\a
\g}-T_{\g}g_{\a \b}) -\frac{1}{6} \epsilon_{\a\b\g\mu} S^{\mu} +
q_{\a\b\g}. \er

It is also useful to write the scalar curvature in terms of these
irreducible components; \br \label{curvaturetor} \widetilde{R} =
R-2 \nabla_{\a}T^{\a}- \frac{4}{3} T^{\a} T_{\a} +
\frac{1}{2}q_{\a\b\g}q^{\a\b\g}+ \frac{1}{24} S^{\a} S_{\a}. \er
The covariant derivative $\nabla_{\a}$ and the Riemannian
curvature tensor $R_{\, \b\g\d}^{\a}$ are obtained from the
symmetric connection $\G_{\,
\b\g}^{\a}=\frac{1}{2}g^{\a\t}(\pa_{\b}g_{\t\g}+ \pa_{\g}
g_{\t\b}-\pa_{\t}g_{\b\g})$.

The general non-minimal action for the scalar field coupled to
metric and torsion is given by \cite{shapiro}
\begin{eqnarray}
\label{action}
S = \int \sqrt{-g}\,  \frac{1}{2} \left[ g^{\mu\nu}
\partial_{\mu} \phi \; \partial_{\nu} \phi - m^2 \phi^2 + \sum_{i=1}^{5}\xi_{i}P_{i} \phi^2 \right] d^4 x \;\;\; .
\end{eqnarray}

We have the following structures: $P_{1}= R $ (the Riemannian
curvature scalar), \,\, $P_{2}= \nabla_{\a}T^{\a},\,\, P_{3}=
T^{\a} T_{\a}$,\,\, $P_{4}= S^{\a} S_{\a}$,\,\,
$P_{5}=q_{\a\b\g}q^{\a\b\g}$. Therefore, there are five
non-minimal parameters $\xi_{1},...,\xi_{5}$. In the torsionless
case the only non-minimal term $\xi_{1} R$ must be
considered\footnote{ We consider the approach in which torsion
mass is dominating over the possible kinetic terms. In the case of
fermion coupled to torsion this assumption is sufficient to
provide the contact spin-spin interactions \cite{Hammond,
shapiro}.}.

Our aim is to find the semi-classical Hamiltonian of the massive
scalar theory coupled to external gravity and on-shell torsion
fields with the general couplings $\xi_{i}$ (\ref{action}) in the
context of the exact Foldy-Wouthuysen transformation\cite{prd}.
Then, making use of the decomposition (\ref{decom}), the equations
of motion for the torsion tensor can be written as the equations
of motion for the relevant components $T_{\\},\, S_{\a}$ and
$q_{\a\b\g}$, \br T_{\a} = \frac{\xi_{2}}{\xi_{3}}
\frac{\nabla_{\a}\phi}{\phi},\,\,\,\,\, S_{\a}\,=\, q_{\a\b\g}=0.
\er

This relation is in accordance with the result that scalar fields produce torsion
only in non-minimally coupled theories. Then, the torsion is related to the gradient
of the field \cite{German, shapiro} (in general,  spin is considered as the source of
torsion, however, the emergence of torsion in other contexts has been discussed in
e.g. \cite{Capozziello}).

Substituting these expressions into (\ref{action}) one gets the
on-shell action
\begin{eqnarray}
\label{action1}
S = \int \sqrt{-g}\,  \frac{1}{2} \left[ g^{\mu\nu}
\partial_{\mu} \phi \; \partial_{\nu} \phi - \hat{m}^2 \phi^2 + \hat{\lambda} \phi^2 R \right] d^4
x,
\end{eqnarray}
where \br \label{lcoup}
\hat{m}^2=\frac{1}{1-\frac{\xi_{2}^2}{\xi_{3}}} m^2,\,\,\,\,\,
\hat{\lambda} =\frac{\xi_{1}}{1-\frac{\xi_{2}^2}{\xi_{3}}}. \er

The effect of torsion coupled to scalar and metric fields are
encoded in the $\xi_{2}$ and $\xi_{3}$
 parameters dependence of the re-defined coupling $\hat{\l}$ and
 mass $\hat{m}$ parameters of the action (\ref{action1}). The mass redefinition due
to the effect of torsion trace has also been reported in
\cite{zecca}.

Notice that an attempt to directly generalize the Einstein-Cartan
theory coupled to a scalar reduces the number of free parameters
$\xi_{i}$ to just one parameter. To observe this, write the
``minimal'' action \br \label{actionmi} S = \int \sqrt{-g}\,
\frac{1}{2} \left[ g^{\mu\nu}
\partial_{\mu} \phi \; \partial_{\nu} \phi - m^2 \phi^2 + \l \widetilde{R}\, \phi^2 \right] d^4 x\, , \;\;\;
\er where the expression (\ref{curvaturetor}) for the curvature in
the space with torsion  must be used. Therefore, comparing
(\ref{action}) and (\ref{actionmi}) we get \br \label{param}
\xi_{1}= \l\,,\,\,\,\xi_{2}= -2 \l ,\,\,\,\xi_{3}= -\frac{4}{3}
\l, \,\,\,\,\xi_{4}= \frac{1}{2} \l,\,\,\,\xi_{5}= \frac{1}{24} \l
\er

In particular, using the relations (\ref{param}) for the redefined
$\hat{\l}$ and $\hat{m}$ parameters one gets \br \label{lcoup1}
\hat{m}^2=\frac{1}{1+ 3 \l} m^2,\,\,\,\,\, \hat{\lambda}
=\frac{\l}{1+3 \l}. \er

Observe that the usual value in the torsionless case
$\hat{\lambda} =\xi_{1}=1/6$ (for massless scalar) corresponds to
$\xi_{1}=\l= 1/3$ when torsion is considered. The shifting of the
conformal value from $1/6$ to $ 1/3$ is due to the non-trivial
transformation of torsion under conformal symmetry \cite{shapiro,
helayel, park}.

\section{Exact Foldy-Wouthuysen transformation}
\label{sec:efw}

In order to find the non-relativistic Hamiltonian of the system we
will use the procedure developed in \cite{prd} for a scalar
 coupled non-minimally to gravity. This problem has been addressed for
a real spin-0 particle coupled to the static metrics
\begin{eqnarray}
\label{metric}
 ds^2 = V^2 dt^2 - W^2 d {\bf{x}}^2,
\end{eqnarray}
where
$V=V({\bf{x}})$ and $W=W({\bf{x}})$.

The treatment uses the properties of a pseudo-Hermitian
Hamiltonian \cite{Mostafazadeh1, Mostafazadeh2, bender} which
appears when the Klein-Gordon equation for the model
(\ref{action1}) is written in the two-component Schr\"odinger
formulation \cite{Feshbach}\br \label{schrodinger1} i \dot{\Phi} =
{\cal{H}} \Phi \;\;\;, \er with the Hamiltonian given by \br
\label{hamiltonian} {\cal{H}} = \frac{\hat{m}}{2} \zeta^T - \zeta
\theta \;\;\; , \er where \br \Phi = \left(
\begin{array}{c}
  \phi_1  \\
  \phi_2
\end{array}\right)  \;\;\;, \;\;\;
\zeta= \left(
\begin{array}{cc}
  1 & 1 \\
 -1 & -1
\end{array}\right)
\er and the operator $\theta$ is defined by \br \theta \equiv \frac{F^2}{2\hat{m}}
\nabla^2 - \frac{F^2}{2\hat{m}} {\boldmath{\nabla}} \ln (VW) \cdot
{\boldmath{\nabla}} - \frac{\hat{m}}{2} V^2 - \frac{\lambda}{2 \hat{m}}V^2
R,\,\,\,\,\,\,\, F^2\,\equiv \, \frac{V^2}{W^2}.\er

Notice that the Hamiltonian satisfies the pseudo-Hermiticity
property ${\cal{H}}^{\dagger}=\s_{3}{\cal{H}}\s_{3}$,\, $\s_{3}$
being the diagonal Pauli matrix diag$(1, -1)$.  Here we simply
quote the exact Foldy's Hamiltonian \cite{prd} \br
\label{foldy1}{\cal{H}}'' = (-2\hat{m} \theta^{'})^{1/2} \s_{3},
\er where

\br \nonumber \theta' &=& - \frac{\hat{m}}{2} V^2 - \frac{1}{2\hat{m}}F {\hat{p}}^2 F
+ \frac{1}{8\hat{m}} {\boldmath{\nabla}} F \cdot {\boldmath{\nabla}} F +
{\cal{D}}_{\hat{\lambda}}(V, W),\,\,\,\, \theta'=f \theta
f^{-1},\\
\nonumber
f&=& V^{-1/2} W^{3/2}, \,\,\,\, \hat{p}=-i\nabla\\
  {\cal{D}}_{\hat{\lambda}}(V, W) &=&
\hat{\lambda} [ (\frac{1}{2\hat{\lambda}}-2)\frac{V}{W^2}\nabla^2 V-
2\frac{V}{W^3}\nabla V. \nabla W + (\frac{1}{2\hat{\lambda}}-4) \frac{V^2}{W^3}
\nabla^2 W + 2\frac{V^2}{W^4} (\nabla W)^2 ]. \nonumber\\ \label{darwin}
 \er

The Eq. (\ref{foldy1}) has been obtained in two steps. First, the operator $\theta'$
is constructed demanding it to be hermitian with respect to the usual flat space
measure. Second, it is a simple observation that ${\cal H}^2$ is diagonal, so, taking
the square-root of this operator and conveniently diagonalizing the $2\mbox{x}2$
identity matrix provides (\ref{foldy1}). However,
 as pointed out in \cite{silenkoprd} a simple diagonalization procedure of the
 Hamiltonian may be nonequivalent to the Foldy-Wouthuysen  transformation. As it is
 well known the FWT
 provides the correct physical interpretation of Klein-Gordon equation written in the
 form (\ref{schrodinger1})-(\ref{hamiltonian}) \cite{bjor}. The Hamiltonian
  (\ref{foldy1}) is exactly the same as
 the one recently proposed in Eq. (83) of Ref. \cite{Mostafazadeh1}, in which a
 rigorous Hilbert space construction
based on the solutions of a Klein-Gordon-type field equation is considered for
$t-$independent $\theta$ operator. As pointed out in \cite{Mostafazadeh1} the Eq.
(\ref{foldy1}) is the Foldy-Wouthuysen Hamiltonian in the Schr\"odinger picture of
the first quantized scalar field theory. According to the discussions above we will
consider ${\cal H}''$ in  (\ref{foldy1}) as the true FW Hamiltonian.

 The quasirelativistic Hamiltonian and the first order terms in the $1/\hat{m}$ expansion
of (\ref{foldy1}) becomes
\begin{equation}
\label{quasire1} {\cal{H}}'' \approx  \Big[ \hat{m} V + \frac{1}{4
\hat{m}} \left( W^{-1} {\hat{p}}^2 F + F {\hat{p}}^2 W^{-1}
\right) - \frac{1}{8 \hat{m} V} {\bf{\nabla}} F \cdot
{\bf{\nabla}}F + \frac{1}{2 \hat{m} V}
{\cal{D}}_{\hat{\lambda}}(V, W)\Big] \s_{3}  \;\;\; .
\end{equation}

The so-called ``gravitational Darwin term" (GDT) is given by
$\frac{1}{2\hat{m} V} {\cal{D}}_{\hat{\lambda}}(V, W)$\,
\cite{prd}. A remarkable fact is that one can rewrite
(\ref{darwin}) as
\begin{eqnarray}
\label{darwin1}
{\cal{D}}_{\hat{\lambda}}(V, W)\, \equiv \,  \hat{\lambda} F \nabla^{2} F + 3 (\frac{1}{6}-\hat{\lambda}) \frac{F}{W}\{\nabla^{2} V +  F \nabla^{2} W \}.
\end{eqnarray}

Notice that the last term in (\ref{darwin1}) inside brackets does
not contribute if $\hat{\lambda}=1/6$ (for the ``minimal" action
(\ref{actionmi}) and according to the Eq. (\ref{lcoup1}) and the
discussion below it, this corresponds to $\xi_{1}=1/3$) providing
a simple form for the Darwin term as discussed in some detail in
Ref \cite{mpla}. The last term in (\ref{darwin1}) for general
$\hat{\lambda}$, of course, does not give a vanishing contribution
and then provides a complicated Darwin term.

\section{The semi-classical approximation for the Hamiltonian}
\label{sec:hamiltonian}

The external gravitational field is assumed to be weak, then a
Newtonian approximation will be sufficient. Thus, far from the
source the solution of the Einstein equation for a point particle
of mass $M$ located at $r=0$ can be taken as
\begin{eqnarray}
\label{g1}
g_{00}\,\approx\, 1-\frac{2MG}{r},\\
\label{g2} g_{11}\,=\,
g_{22}\,=\,g_{33}\,\approx\,-1-\frac{2MG}{r}.
\end{eqnarray}

From (\ref{g1}) and (\ref{g2}) we get immediately

\begin{eqnarray}
\label{v1}
V\,\approx\,1-\frac{MG}{r},\,\,\,\,\,\,\,
W\,\approx\,1+\frac{MG}{r}\\
\label{f1}
\mbox{and}\,\,\,\,\,\,  F\,\approx\, 1- 2 \frac{MG}{r}.
\end{eqnarray}

Inserting (\ref{v1}) and (\ref{f1}) into  (\ref{quasire1}) we
obtain the non-relativistic FW Hamiltonian
\begin{equation}
\label{f12} {\cal{H}}'' =\left[ \hat{m}+ \hat{m} \; {\bf g} \cdot
{\bf x} + \frac{\hat{{\bf p}}^2}{2\hat{m}} + \frac{3}{2\hat{m}}
\hat{{\bf{p}}} \cdot({\bf g} \cdot {\bf x}) \hat{{\bf{p}}} -
\frac{4\pi GM}{m} \lambda_{\xi_{i}} \,  \,
\d^{3}(\vec{\bf{r}})\right] \s_{3}, \;\;\;
\end{equation}
where $ {\bf g} = - GM \frac{{\bf r}}{r^3}$ and $\lambda_{\xi_{i}}
\equiv \xi_{1} / \sqrt{1-\frac{\xi_{2}^2}{\xi_{3}}}$.

The `mass' $\hat{m}$ is the redefined parameter in (\ref{lcoup}).
It is  known that a spinless particle in an external torsion field
undergoes a shift of its mass \cite{shapiro, zecca}.

The first three terms in (\ref{f12}) represent the rest energy,
the gravitational potential, the non-relativistic kinetic term,
respectively, while the fourth term is the first relativistic
correction for the gravitational potential. The last term
proportional to $\nabla^2 \frac{1}{r} \sim \d^{3}(\vec{\bf{r}})$
in (\ref{f12}) can be interpreted as a 'gravitational Darwin term'
in analogy to the usual electric Darwin term $\nabla . \bf{E}$
arising from `zitterbewegung' (the particle's coordinate is
`smeared out' over a length $\approx \hbar/mc$).

In order to get (\ref{f12}) we have used the fact that for the special type of
metrics (\ref{metric}) and in the approximation (\ref{v1})-(\ref{f1}), the last term
of (\ref{darwin1}) containing the brackets gives a negligible contribution for
$\hat{\l}\neq \frac{1}{6}$. So, the GDT arises only from the $\hat{\l} F \nabla^2 F$
sector of (\ref{darwin1}) \br \label{delta} \frac{\hat{\lambda}}{2 \hat{m} V}
{\cal{D}}_{\hat{\lambda}}(V, W) \approx  -\frac{4\pi GM}{m} \lambda_{\xi_{i}} \,
\d^{3} ({\bf r}), \er where in the denominator of the right hand side one has  the
scalar mass $m$ parameter of action (\ref{action}). For the case (\ref{lcoup1}) the
relationship $\lambda_{\xi_{i}} = \xi / \sqrt{1+ 3\xi}$\, holds.

The GDT without torsion presented in  \cite{mpla}, in the case under consideration
here gets redefined by a coefficient depending on the extra $\xi_{2, \, 3}$
couplings. Moreover, the GDT (\ref{delta}) is the analog to the one obtained for the
torsionless case in the effective field theory approach \cite{flachi} where a $\l$
dependent term $\lambda_{\xi_{i}}(\l)$ has been obtained as the next to leading
correction to the Newtonian potential at tree level due to the non-minimal coupling.

On the other hand, this contribution to the static gravitational
potential is similar to the one obtained in quadratic gravity.
Quadratic gravity is constructed by the addition of non-linear
terms to the curvature in  the Einstein-Hilbert action. The
coupling parameters of these terms in the Lagrangian must be
determined by experiments. Experimental constraints on these
parameters can be set out from the dispersive character of the
bending of light in higher derivative gravity and based on the
fact that rainbow effect is currently undetectable \cite{prd1} or
from sub-millimeter tests of the inverse square law \cite{long,
hoyle}. To gain insight into the nature of the term (\ref{delta})
let us write the potential which follows from quadratic gravity
\cite{stelle, azeredo} \br \label{poten1} V(r)\,=\, -\frac{G M
m}{r} (1+ \frac{1}{3} e^{-m_{0} r}-\frac{4}{3} e^{-m_{1} r}), \er
where $m_{0}=\sqrt{\frac{1}{\kappa^2 (3 \a + \b)}}$ and
$m_{1}=\sqrt{-2/ (\kappa^2 \b)}$, $\a$ and $\b$ are dimensionless
parameters, and $\kappa^2 = 32\pi G$ is the Einstein's constant.
The $\a$ and $\b$ are the parameters in the $\a R^2 +\b
R_{\mu\nu}^2$ terms of the higher order gravity
\footnote{Regarding the potentials described above, let us point
out that also a Yukawa type potential has been reported for the
static limit of the vector component of torsion in Eq.
(\ref{decom}) in the framework of transposition invariance
formulation (see \cite{HammondRPP} and references therein).}.

Experimental bounds on the parameter $\a$ and $\b$ given in Refs.
\cite{prd1, long, hoyle} are \br \a\,,|\b| \, <\, 10^{60}, \er
which improves the earlier bound $10^{74}$ of Stelle \cite{stelle,
azeredo}.

In the limit $\a\, , \b\, \rightarrow 0$, as is usually more
appropriate for a perturbation in an effective field theory, the
potential (\ref{poten1}) at first order becomes \cite{Donoghue,
donoghue1} \br V(r)\,=\, -G M m \( \frac{1}{r} - 128 \pi^2 G (\a +
\b) \d^{3}(\bf{\vec{r}}) \).\label{pot11}\er

The above limiting procedure provides the low energy potential.
Then the $R^2$ terms give rise to a very weak and short-ranged
contribution to the Newtonian potential.

The last term contribution obtained in (\ref{f12}) due to the
non-minimal couplings has the same structure as the delta function
term in (\ref{pot11}). Therefore, since the torsion field effect
on the scalar particle manifests itself by redefining the GDT
coefficient \footnote{Regarding this point, the authors of Ref.
\cite{Aprea} considered propagating torsion effects on test
particles and observed that the non-relativistic limit of the
auto-parallel equation shows that the ``force" due to the torsion
potential is manifested in the same way as the gravitational one.
Moreover, for static and weak potentials of the trace part of the
torsion $\vp$ and the gravitational field $h_{00}$ they showed
that in the non-relativistic limit the potentials obey a Poisson
partial differential equation and concluded that due to their
similar effects on the test particle it is impossible to
distinguish their effects unless the source and the initial
condition for the
 torsion field is supplied; thus, the smallness of the intensity of the torsion
forces, makes their detection even more difficult.}, following
similar arguments used in the quadratic gravity case and in the
effective field theory approach to dealing with the non-minimal
coupling \cite{flachi}, one may conclude that it is not possible
to measure the non-minimal coupling $\l_{\xi_{i}}$ in the region
of energy/curvature where the non-relativistic approximation is
valid. Then, the effective field theory of gravity is not affected
by the presence of the non-minimal coupling terms $\xi_{i}$ in
(\ref{action}).

\section*{Acknowledgments}

 AA thanks FAPERJ-Brazil and CNPq-Brasil for partial support. HB thanks the
 Departamento de Matem\'atica-ICET, UFMT for hospitality, R. Ochoa for
calling his attention to an earlier reference and S.S. Costa for reading the
manuscript, and CNPq-FAPEMAT for financial support.

\end{document}